\documentclass[unnumsec,webpdf,contemporary,large]{brsl-template}%

\usepackage{microtype}
\usepackage{cleveref}

\usepackage{paratype} 
\usepackage[T1]{fontenc}

\graphicspath{{Fig/}}

\theoremstyle{thmstyleone}%
\theoremstyle{thmstyletwo}%
\theoremstyle{thmstylethree}%

\begin{document}

\journaltitle{BRSL Research Preprint}
\copyrightyear{2024}
\pubyear{May 2024}
\appnotes{\hspace{1mm}}

\firstpage{1}


\title[Open-Source Assessments of AI Capabilities]{Open-Source Assessments of AI Capabilities: \Large{The Proliferation of AI Analysis Tools, Replicating Competitor Models, and the \textit{Zhousidun} Dataset}}

\author[1,2,$\ast$]{Ritwik Gupta\ORCID{0000-0001-7608-3832}}
\author[1]{Leah Walker}
\author[1]{Eli Glickman}
\author[1]{Raine Koizumi}
\author[1]{Sarthak Bhatnagar}
\author[1]{Andrew W. Reddie\ORCID{0000-0003-3231-8307}}

\authormark{Ritwik Gupta et al.}

\address[1]{\orgdiv{Berkeley Risk and Security Lab}, \orgname{University of California, Berkeley}}
\address[2]{\orgdiv{Berkeley AI Research Lab}, \orgname{University of California, Berkeley}}

\corresp[$\ast$]{Corresponding author: \href{ritwikgupta@berkeley.edu}{ritwikgupta@berkeley.edu}}



\abstract{
The integration of artificial intelligence (AI) into military capabilities has become a norm for major military power across the globe. Understanding how these AI models operate is essential for maintaining strategic advantages and ensuring security. This paper demonstrates an open-source methodology for analyzing military AI models through a detailed examination of the \textit{Zhousidun} dataset, a Chinese-originated dataset that exhaustively labels critical components on American and Allied destroyers. By demonstrating the replication of a state-of-the-art computer vision model on this dataset, we illustrate how open-source tools can be leveraged to assess and understand key military AI capabilities. This methodology offers a robust framework for evaluating the performance and potential of AI-enabled military capabilities, thus enhancing the accuracy and reliability of strategic assessments.
}

\maketitle

\section{Uncovering the \textit{Zhousidun} Database}
In August of 2023, researchers at the Berkeley Risk and Security Lab came across a dataset of Chinese origin named \textit{Zhousidun} (or ``Zeus’s Shield,'' using the Chinese colloquial name for the U.S. Aegis Combat System).\footnote{The full Mandarin name for the Aegis Combat System is \textit{Zhousidun Zhandou Xitong}.} The dataset was made publicly available on Roboflow, a platform where users can share datasets and train machine learning models, likely by ShanghaiTech University.\footnote{We believe that this dataset is most likely from ShanghaiTech’s \href{https://sist.shanghaitech.edu.cn/sist_en/2017/0313/c3902a31619/page.htm}{Visual and Data Intelligence Center (VDI Center)}. The name of the account that posted the dataset to RoboFlow was ``shanghaitech-faxfj.''} The \textit{Zhousidun} dataset is composed of 608 oblique and satellite images of American \textit{Arleigh Burke}-class destroyers and other allied destroyers and frigates (a complete list of which is provided in Appendix \ref{app:vessels}).

\begin{figure*}
    \centering
    \includegraphics[width=\textwidth]{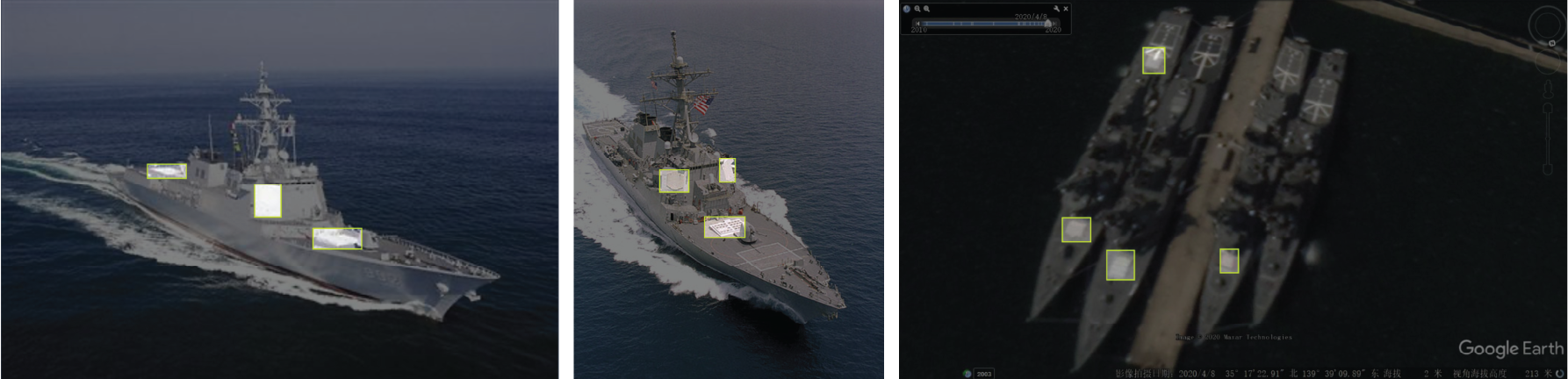}
    \caption{Images from the \textit{Zhousidun} dataset, which feature various military vessels with overlaid bounding boxes on SPY radars and VLS missile-launching components.}
    \label{fig:zhousidun-boxes}
\end{figure*}

The nature of this dataset is highly specific and atypical — academic work in this space is largely focused on whole ship detection, not on specific components of military vessels. In those \textit{Zhousidun} images with \textit{Arleigh Burke}-class destroyers, the ships’ radar systems (which are part of the Aegis Combat System\footnote{An integrated, networked combat system deployed on U.S. Navy destroyers and cruisers for fleet air and missile defense.}), are labeled with bounding boxes on each image (see Figure \ref{fig:zhousidun-boxes}). The images in the dataset are obtained from publicly available sources with some watermarked with media outlet information and the Google Earth logo. Due to the targeted, military nature of the dataset and the likely academic origins of the account sharing it, we suggest that it is likely that this dataset was accidentally published. 

\begin{figure}
    \centering
    \includegraphics[height=2in]{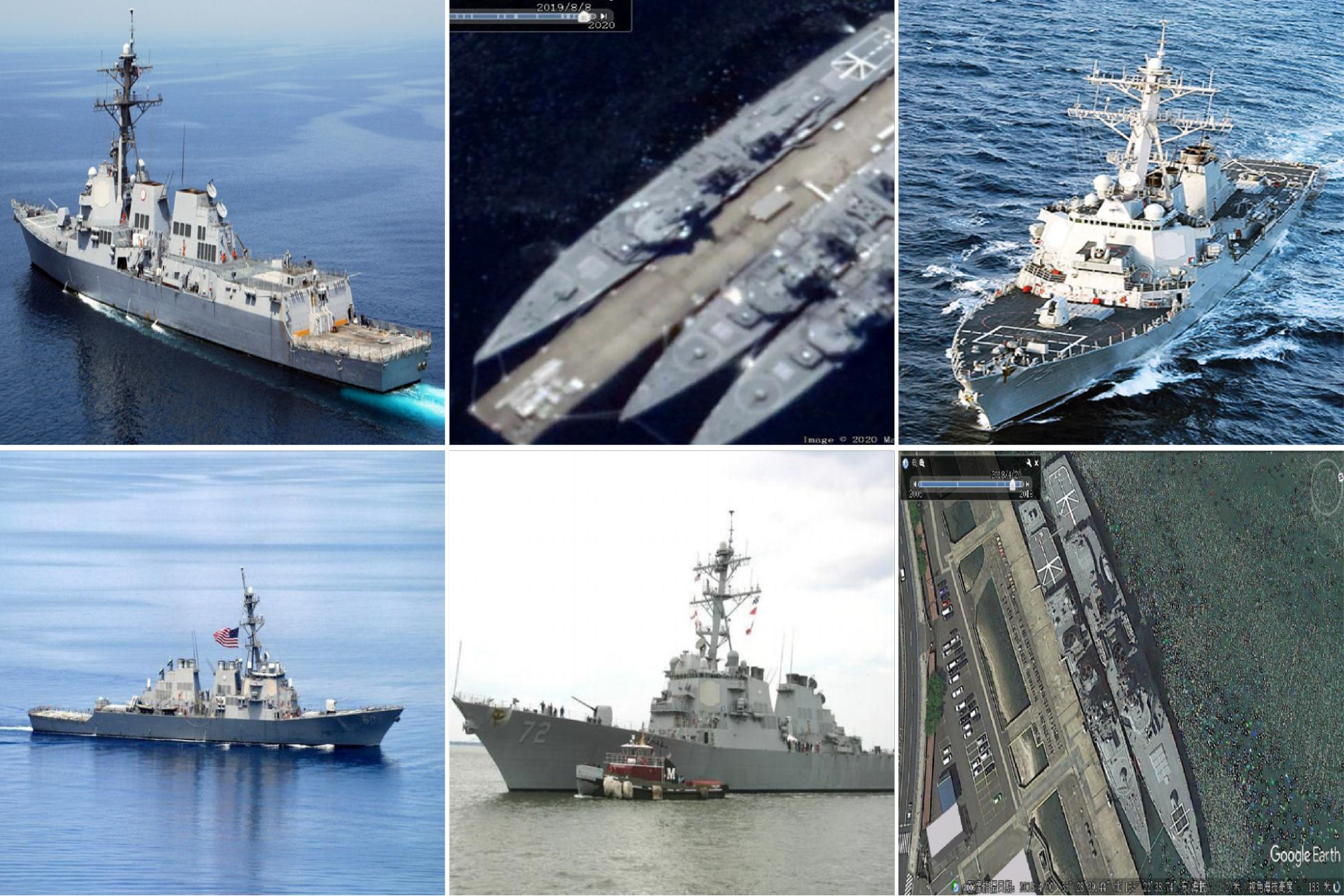}
    \caption{Additional examples of images from the \textit{Zhousidun} dataset.}
    \label{fig:zhousidun}
\end{figure}

In this paper, we demonstrate a quantitative method to assess the AI capabilities of a third-party through the replication of a near-state-of-the-art computer vision model on the \textit{Zhousidun} dataset. This method can be applied to other, similar, databases to further our understanding of competitor models and AI capabilities. We then demonstrate the performance of the trained model on this exposed dataset. Furthermore, we create a synthetic data generation process around a U.S. destroyer from the perspective of a small unnamed aerial system and demonstrate the performance of the trained model on this data. Finally, we hypothesize as to why Chinese affiliates may be using this approach as opposed to others and its potential implications for the U.S. Navy.

\section{The Proliferation and Lifecycle of AI Intelligence Tools}

While the dataset is focused on warships, any conclusions about its intended use must be caveated by the limited information we have about its origins. We cannot say with certainty whether this database was developed by a researcher at a university interested in testing new tools for ship detection, or a navy computer scientist as part of a military project, or even an undergraduate student looking to impress potential employers in the defense sector. That all three of those profiles could fit as creators of this database is a testament to the increasing availability of AI intelligence tools outside of traditional military domains and the reduced skill barrier to using them. In fact, in this scenario, it cannot be assumed that the military scientist would even be the most capable in compiling or training models. Based on the username of the person who released the dataset, it would appear the \textit{Zhousidun} originated from ShanghaiTech University. We can deduce no additional information about the origin of the dataset.

While historically, military R\&D has kept up with, or even led, innovation in the private sector, this is no longer the case. Exquisite military technologies were generally developed in isolation and kept secret, with the know-how and talent to develop these technologies further cloistered, ensuring that the replication of the capability was near impossible externally. The world of artificial intelligence, however, represents a marked exception to this status quo, as AI research and development has been widely democratized. Pioneering artificial intelligence research is, largely, not kept for private profit but is instead almost immediately released into the public domain via publication and paired release of code. The tools, techniques, and procedures to replicate and extend these capabilities are available at the click of a button and the skills required to use them are learnable through the course of a day on YouTube.\footnote{You can see excellent courses such as \href{https://course.fast.ai/}{Practical Deep Learning for Coders} or \href{https://www.youtube.com/watch?v=Bl4Feh_Mjvo&list=PLoROMvodv4rNyWOpJg_Yh4NSqI4Z4vOYy}{Stanford’s CS229 lectures}.}

With the realization that academics, companies, and governments alike are all using the same methods, and often the same publicly released code, replicating the models that are utilized by adversaries becomes almost trivial. Indeed, the differentiating factor between capabilities is increasingly the data used to train the models as well as the pipelines surrounding data acquisition, model verification and validation, and testing \cite{shankarOperationalizingMachineLearning2022}.

Data and methods to source said data have become closely guarded secrets by almost all organizations in the pursuit of ``better'' AI capabilities. It is rare to find data used by organizations for proprietary or restricted AI capabilities online, making replication of these models difficult. Even ``open'' AI models, those that publicly release their model weights such as Meta’s Llama or Google’s Gemma, do not release the underlying data that they were trained upon.\footnote{With the exception of a few new models, such as the Allen Institute’s \href{https://allenai.org/olmo}{OLMo} model.}

When datasets from adversaries or competitors are accidentally revealed, they offer an opportunity for deep, targeted exploration of the capabilities of models trained on those datasets and the limits of the collection capabilities of those adversaries. Additionally, advances in simulation and modeling make it possible to meaningfully replicate data, thus providing a more holistic testing distribution than the original dataset alone. In cases where no dataset is available, simulation and modeling make it possible to bootstrap a testing dataset entirely.

\subsection{Bridging the Gap Between Academia and the Military}
Due to the rapid pace of innovation in the field of AI, the tools and technologies that are used by undergraduates in their introduction to deep learning courses are oftentimes more advanced than what are available to the world’s best militaries. This is in complete opposition to the status quo in which militaries generally have exquisite capabilities not otherwise realized by the world at large.

This gap in capability between sectors underlines the importance of an academia-to-military acquisition and sustainment cycle for the success of military AI efforts. The United States and China are no strangers to the integration of academia and military efforts. In the United States, much research in advanced topics in AI is funded by the Department of Defense via the Defense Advanced Research Projects Agency (DARPA)\footnote{See DARPA’s \href{https://www.darpa.mil/work-with-us/ai-forward}{AI Forward strategy}, examples of which include the DARPA \href{https://www.darpa.mil/news-events/2023-09-25}{ANSR}, \href{https://www.darpa.mil/program/explainable-artificial-intelligence}{XAI}, and \href{https://www.c4isrnet.com/battlefield-tech/space/2022/04/22/darpa-seeks-proposals-on-improving-satellite-imagery-technology/}{FIDDLER}.} or various service research labs. In China, many academic AI efforts are funded by the National Natural Science Foundation and National Key R\&D Programs; these are civilian programs that analysts believe also funnel PLA funds \cite{acharyaChinesePublicAI2019}.

Funded academic AI research is not guaranteed to make it into an operational military capability. Much has already been discussed about the ``valley of death'' in defense acquisitions \cite{landrethDoDValleyDeath2022}, but it is important to underline that while even the leading militaries in the world struggle to transition their R\&D, many countries struggle even more. For example, the leading AI research institutions in Germany rarely transition their research directly to the Germany military or NATO, in part because of viewing AI through mainly economic and societal lenses, rather than military ones \cite{frankeNotSmartEnough2019}.

\subsection{AI Ship Detection}
One area where academic computer vision research has overwhelmingly embraced expanded public access to satellite imagery is that of ship detection and identification. Researchers are able to identify and track objects and vessels at sea by applying computer vision algorithms to satellite imagery.

These techniques have both military and civilian value. On the civilian side, being able to monitor and identify objects at sea from commercial satellite imagery can enable tracking illicit fishing, identifying ships for sanction evasion, and monitoring climate and environmental changes such as tracking coral reefs degradation. Of  course, these very techniques can also be used to identify and track, and even target, naval vessels and other at-sea military equipment. 

To highlight this emerging area of research, and also identify ways in which researchers (in government, in academia, in the military, or elsewhere) may approach a similar dataset, we conducted a literature review of 76 papers related to AI ship detection or similar research. A complete bibliography is provided in Appendix \ref{app:papers}.

Collectively, these papers demonstrate a substantial interest in high-fidelity maritime object detection. Many of these papers incorporate statistical and computer vision-based techniques to detect and classify ships against crowded backgrounds, including in ports and along coastlines. Primarily, synthetic aperture radar (SAR) satellite imagery serves as the foundation for AI-based ship detection literature, followed by high-resolution electro-optical satellite and oblique imagery. Researchers often retrieve imagery from publicly available sources for most datasets, including the European Sentinel-1 SAR constellation and imagery scraped from Google Earth. In some instances, datasets use imagery from hard-to-obtain civilian sources, like the Chinese Gaofen constellation, while others depend on maritime surveillance feeds from drones and cameras.

\begin{figure*}
    \centering
    \includegraphics{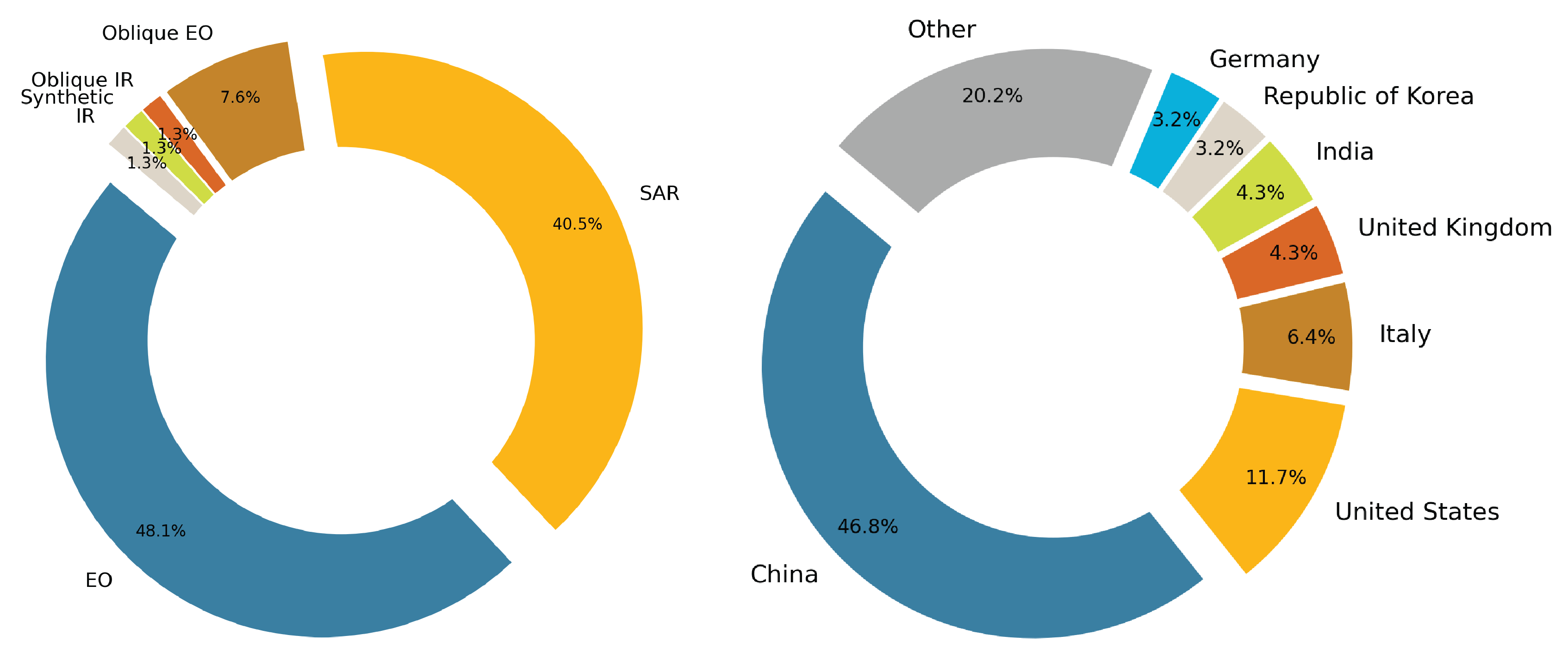}
    \caption{(left) Distribution of types of sensors used in the analyzed papers. (right) Distribution of the countries of origin for the institutions represented in the analyzed papers. A full list of papers is in Appendix \ref{app:papers}.}
    \label{fig:paper-distros}
\end{figure*}

The detection of vessels, especially ones aiming to stay hidden on a vast ocean, remains an unsolved and complicated problem. As outlined in our analysis of datasets above, almost all modern ship detection methods attempt to detect entire classes of ships at once. That is, the entire ship has a bounding box drawn around it. This whole-ship detection and classification approach results in models that are unable to properly differentiate between different types of ships. Often this is due to a limitation in deep learning methods to learn salient features that distinguish a specific kind of ship.

\section{Replicating Models on \textit{Zhousidun}}

\textit{Zhousidun} provided the opportunity for us to demonstrate what we believe is a useful approach to testing and assessing competitor datasets, and by extension, models. In sum, we utilized a widely-used, near-state-of-the-art object detection model to analyze the effectiveness of a model trained on \textit{Zhousidun} on the real-world task of detecting Aegis combat systems in the wild. We perform external validation of this model on synthetic data generated from a custom-built 3D scene of an \textit{Arleigh Burke}-class destroyer. We then demonstrate the limitations of the \textit{Zhousidun} dataset, and therefore, of datasets scraped from the public internet, and discuss how such models might be used in the context of \textit{Zhousidun}’s focus on U.S. and allied naval vessels.

\subsection{Data}
The \textit{Zhousidun} dataset is composed of 608 images of military naval vessels, largely destroyers, that are equipped with the Aegis combat system. These images are from ground-based, oblique optical sensors and optical satellite imagery. Bounding boxes have been drawn around SPY radars on the superstructure, one on port and one on starboard, as well as around the vertical launching systems towards the bow and towards the stern of the ship.

The images comprising \textit{Zhousidun} appear to have been collected from the open Internet due to the inclusion of various watermarks in the images, as well as the unconstrained quality and geometries of the collected images themselves. Indeed, we were able to locate the original sources for the majority of the oblique images in the dataset from a variety of media and military outlets. The satellite images are all low-quality screenshots from Google Earth. Scraping satellite imagery from Google Earth is a common practice in widely-used Chinese academic datasets such as MLRSNet \cite{qiMLRSNetMultilabelHigh2020}, AID \cite{xiaAIDBenchmarkData2017}, WHU-RS19 \cite{daiSatelliteImageClassification2011}, and others.

Together, this means that \textit{Zhousidun} is not a representative dataset of the sensors and images collected by the PLAN. However, bootstrapping powerful object detectors from low-quality, web-scraped imagery is effective and a popular approach in the machine learning community with the advent of webly supervised \cite{chenWeblySupervisedLearning2015} and self-supervised learning techniques.

To ensure that our analysis is representative of realistic, high-quality data of the kind that can be obtained from modern optical sensors, we build a synthetic scene of a static \textit{Arleigh Burke}-class destroyer in the middle of an ocean in moderate sea conditions (Beaufort force 4). Images are collected in a half-dome around the vessel representative of imagery captured from a small, unmanned aerial system or a high-resolution optical satellite. These images are solely used for evaluation.

\begin{figure}
    \centering
    \includegraphics[width=\columnwidth]{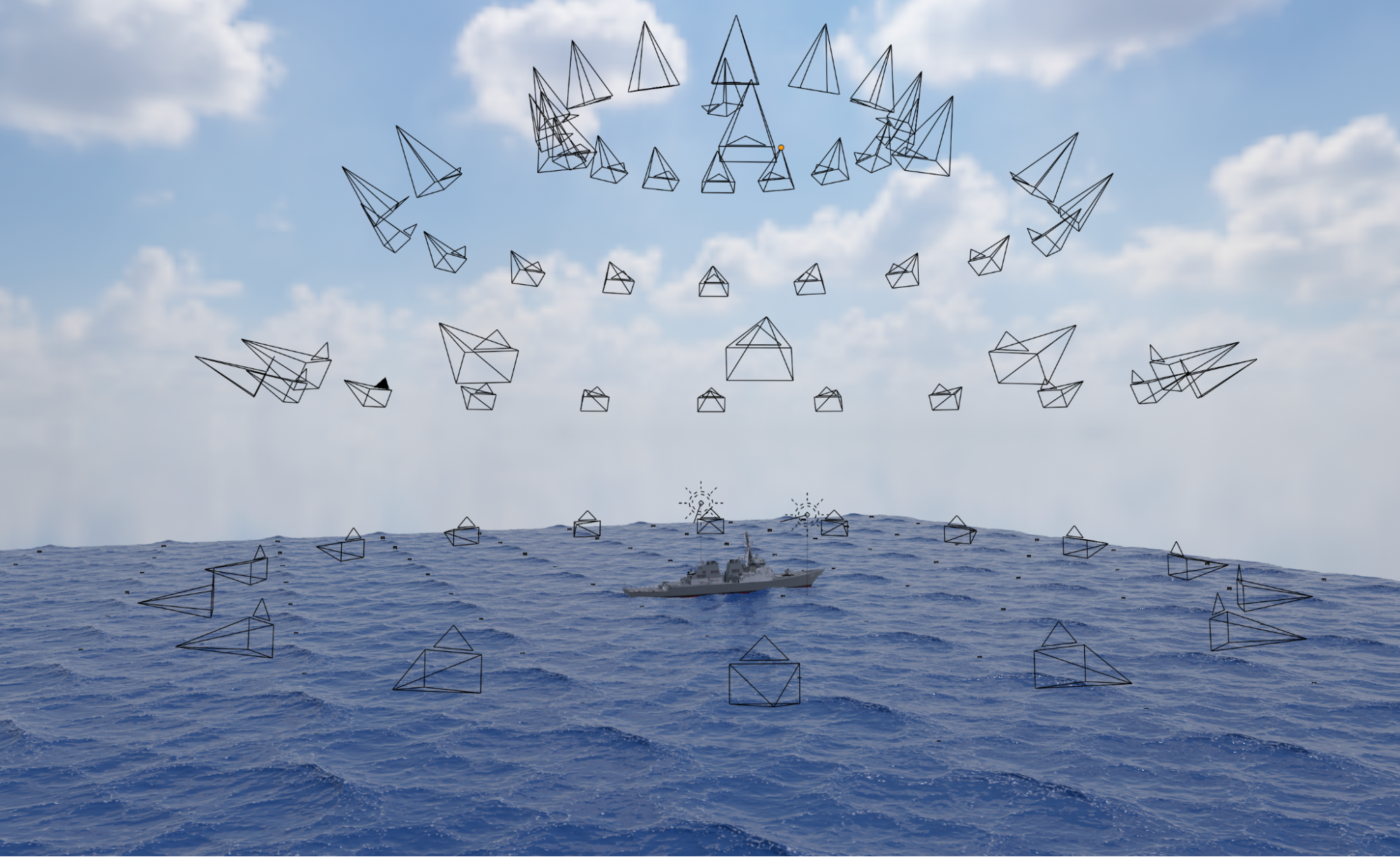}
    \caption{A synthetic scene of the USS \textit{Arleigh Burke} with positions of captured images.}
    \label{fig:blender}
\end{figure}

Appendix \ref{app:data-model} contains further details about how the data is processed for use in a machine learning pipeline.

\subsection{Modeling}
There are a variety of popular, competitive deep learning frameworks for object detection available for use in the open source that can aid in the creation of object detection models from the \textit{Zhousidun} dataset. These frameworks, such as Meta’s Detectron2\footnote{\href{https://github.com/facebookresearch/detectron2}{Detectron2} is a platform for object detection, segmentation and other visual recognition tasks.}, China’s PaddleDetection\footnote{\href{https://github.com/PaddlePaddle/PaddleDetection}{PaddleDetection: Object Detection toolkit based on PaddlePaddle}.}, and others implement near-state-of-the-art models for object detection with an easy-to-use and easy-to-extend interface.

In computer vision research, the mmdetection\footnote{\href{https://github.com/open-mmlab/mmdetection}{GitHub - open-mmlab/mmdetection: OpenMMLab Detection Toolbox and Benchmark}} framework is commonly used due to the extensibility it provides. In industry, one of the most adopted frameworks for object detection is the Ultralytics\footnote{\href{https://github.com/ultralytics/ultralytics}{https://github.com/ultralytics/ultralytics}.} object detection framework. For example, the YOLO family of models in the Ultralytics library is used broadly. YOLO is a single-stage CNN for object detection that is fast, memory efficient, and accurate, all highly desirable characteristics for deployment on edge devices such as small unmanned aerial vehicles (sUAS).

Recently, vision transformers have beat models such as YOLO for object detection on almost all popular object detection benchmarks, such as COCO \cite{linMicrosoftCOCOCommon2014}, by at least 17\% on mAP (a metric described in depth later in this report).\footnote{\href{https://paperswithcode.com/sota/object-detection-on-coco}{COCO test-dev Benchmark (Object Detection) | Papers With Code}.} However, these other methods are more difficult to train and are not optimized for use on compute-constrained platforms such as sUAS. For that reason, Ultralytics’ implementation of YOLO remains the dominant method used by industry today.

For our analysis of the \textit{Zhousidun} dataset, we use the YOLOv8 model implemented by the Ultralytics library. We discussed earlier that cutting-edge machine learning models are developed in the open and are distributed widely immediately; therefore, it is unlikely that any state actor possesses better machine learning models than what exists publicly, this choice is a good representation of what the PLA would be using for their object detection needs.

\subsection{Methodology}
To conduct our analysis, we use the train and test sets of the \textit{Zhousidun} dataset as provided by the dataset uploaders. We train the YOLOv8-large (YOLOv8l) detector on the train split of the dataset.\footnote{Code and weights are made available on \href{https://github.com/BerkeleyRisk/Zhousidun}{BRSL’s GitHub repo}.} Further details about the model configuration are provided in Appendix \ref{app:data-model}.

In this evaluation, we aimed to understand how well a model trained on publicly-available imagery can perform in real-world identification targeting situations. We evaluate the performance of the model through a metric called “mean average precision” (mAP). The mAP metric tells you how “good” a model is at not only finding all of the right objects (a metric called “recall”) but also ensuring that it only gets the right objects (measured through “precision”). mAP ranges between 0 and 1; a higher mAP score indicates that the model is more accurate and misses few objects. We evaluate mAP an intersection over union (IoU) of 0.50, i.e., a predicted bounding box is only correct if it overlaps at least 50\% of the true bounding box.

\begin{figure}[h]
    \centering
    \includegraphics[width=\columnwidth]{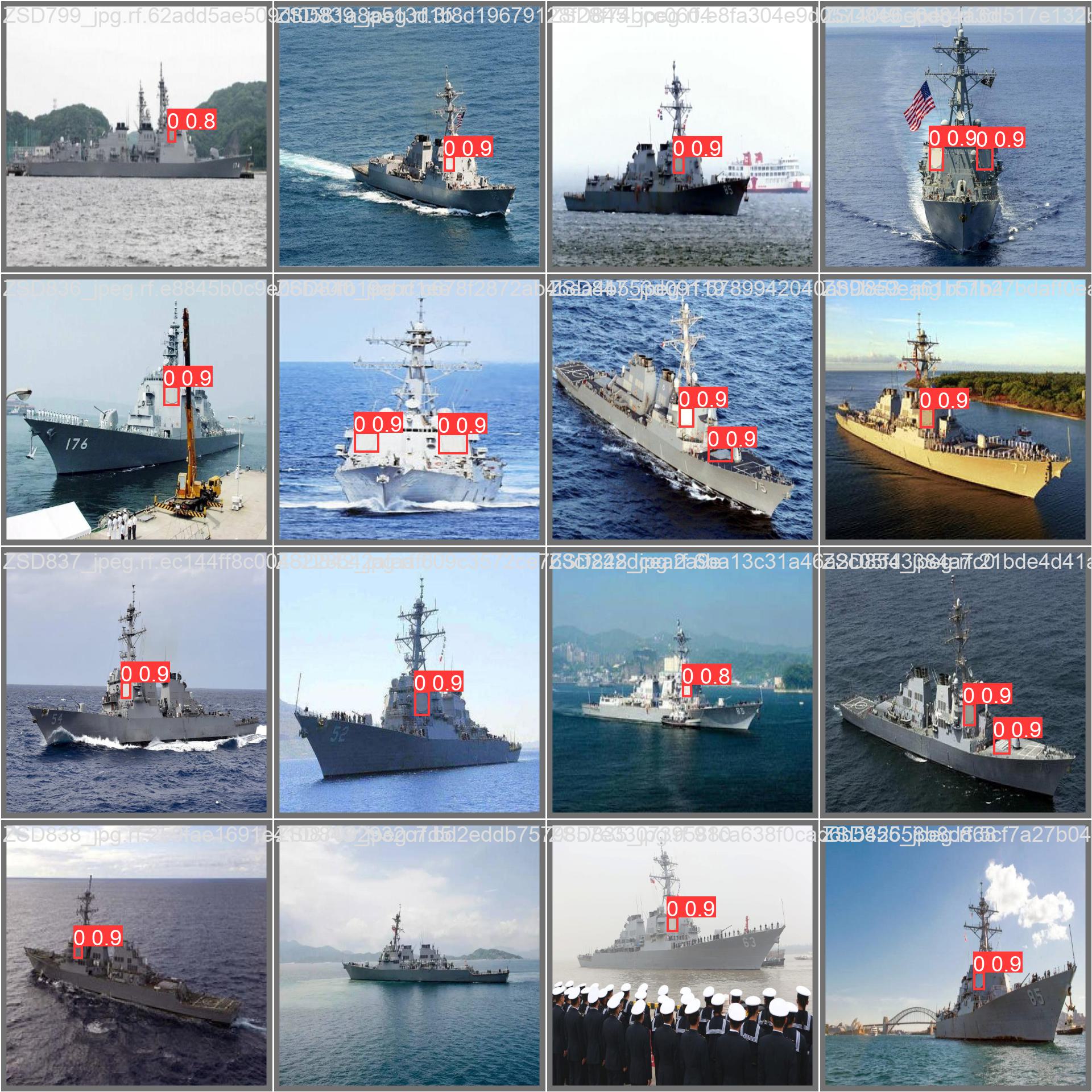}
    \caption{Predictions on the \textit{Zhousidun} test set.}
    \label{fig:pred-zhousidun}
\end{figure}

Evaluated simply on the test set of \textit{Zhousidun}, the trained model achieves a mAP (at 0.50 intersection over union (IoU)) of 0.926. In many targeting workflows, this mAP is considered to be quite good. These results are visualized in Figure \ref{fig:pred-zhousidun}.

However, the test set of \textit{Zhousidun} is not a realistic testing target. It represents an artificial set of images collected from an unrealistic source such as Google Images and does not mirror what the real-world performance of a model trained on this publicly-sourced data would look like.

\begin{figure}
    \centering
    \includegraphics[width=\columnwidth]{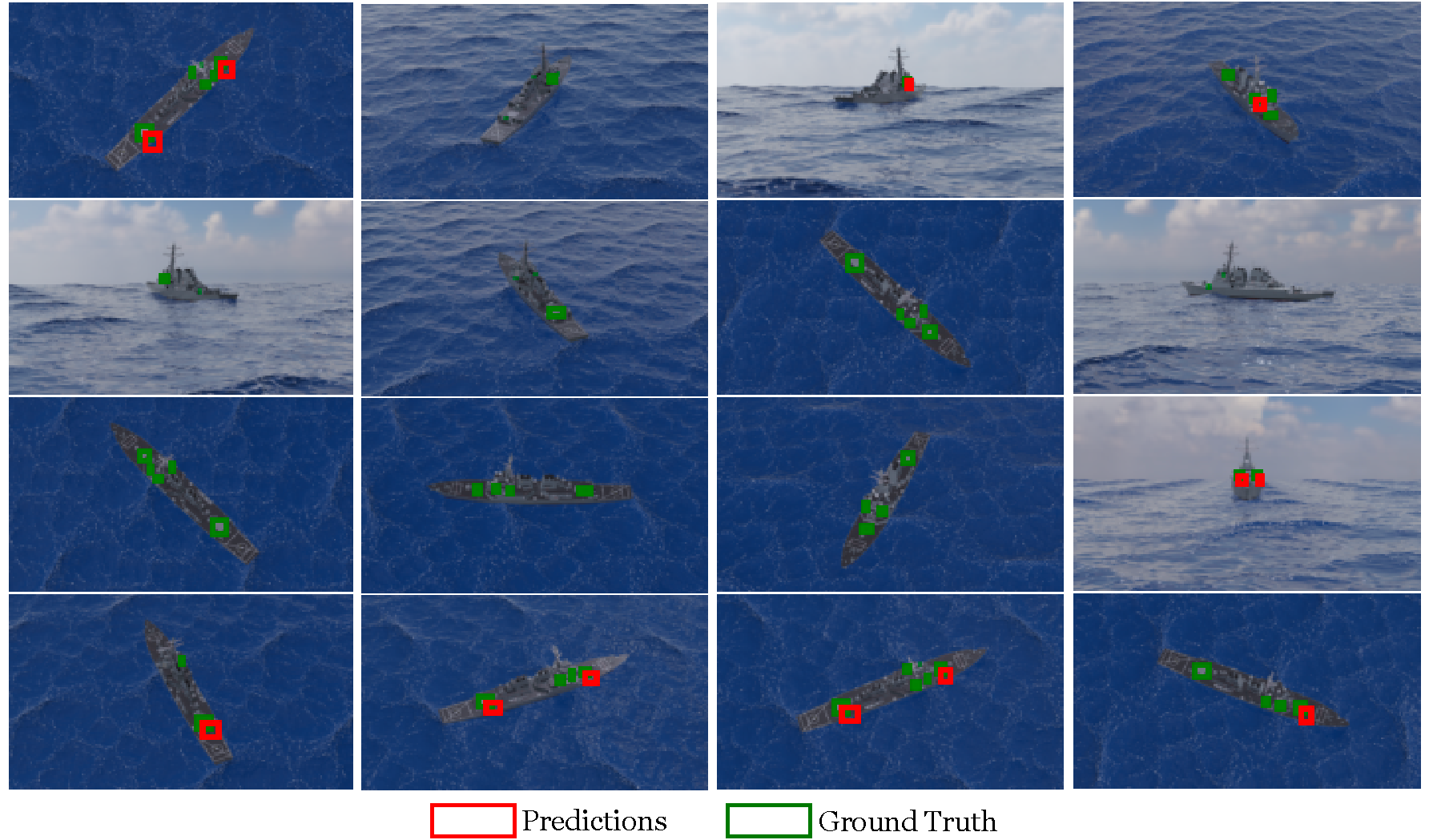}
    \caption{Predictions on the synthetic \textit{Arleigh Burke}-class destroyer dataset from a trained YOLOv8l model on the \textit{Zhousidun} dataset.}
    \label{fig:pred-blender}
\end{figure}

To assess the out-of-distribution performance of this trained model, we generated a synthetic 3D scene of an \textit{Arleigh Burke}-class destroyer on the open ocean. We utilize the same evaluation methodology as above for these synthetic images. Specifically, we evaluated the performance of the model on all geometries of collect, a subset of collects from oblique viewing angles, and a subset of collects from near-nadir angles mimicking satellite imagery. Outputs from this experiment are provided in Figure \ref{fig:pred-blender}.

Across all geometries, the model had a mAP of 0.45 with poor recall (0.26) and decent precision (0.87). This means that the model struggled to initially find the components of the Aegis combat system, but when a component was identified, the resulting detection was precise. When focusing on just oblique geometries, the model’s mAP stayed  relatively constant at 0.49 with comparable recall (0.31) and precision (0.86). However, the model performed much worse on geometries mimicking that of a satellite, achieving only a mAP of 0.34 with extremely poor recall (0.18) but with a drastically higher precision (0.92). This demonstrates that the classifier rarely detects components from near-nadir angles but is largely correct when it does output a detection.

Overall, a model trained on \textit{Zhousidun} has limited targeting capabilities in the real world. It is unlikely that any military would field a model with these performance characteristics. However, it is extremely interesting that training on a small set of unconstrained, publicly available imagery offers such a great starting point to building a robust targeting model. Collecting real images of Allied destroyers with sUAS is not a trivial task and the amount of available data is low. By combining datasets like \textit{Zhousidun} with a small set of real-world images in tandem with modern approaches in pre-training and fine-tuning, we can convert this weak model into a powerful detector.

\subsection{Detection Techniques}
The AI-based ship detection techniques implemented in these papers are largely built on convolutional neural network (CNN) \cite{lecunGradientbasedLearningApplied1998} architectures, which are a class of deep learning algorithms used to perform feature extraction and object detection. Deep learning algorithms are most widely used in visual object recognition. They are organized into many layers, with each layer serving to extract more semantic information from the previous one. CNNs are modeled on the human vision processing system with iterative feature extraction and efficient message-passing interfaces via spatially local connections. \cite{russellArtificialIntelligenceModern2021}

To mitigate challenges associated with noisy images\footnote{``Noisy'' includes everything from poor sensor calibration, degraded image quality due to poor image processing, or images of varying and unreliable resolution characteristics.} and obscured objects,\footnote{Clouds are the most common obscuring factor in satellite imagery.} a number of these techniques couple constant false alarm rate (CFAR) detection with CNNs. \cite{carmanComparisonFixedThreshold2020,zhangNovelFullPolarizationSAR2021,aiAISDataAided2022,benito-ortizDesignEvaluationArtificial2018,aiMultiScaleRotationInvariantHaarLike2019,couchman-crookNovaSARSSTLS142020,leiSRSDDv1HighResolutionSAR2021} Traditional ship detection methods are challenged by small and obscure targets, background noise, and signal interference. CFAR mitigates this by determining the thresholds required to separate the ``true'' targets from these obscure areas within an image. However, this raises the probability of false positives. A number of these papers note that piggybacking CFAR detection with a CNN’s classification capability removes these false positives. Techniques using this architecture demonstrate an increased capacity to identify ships in noisy images and identify smaller, obscured objects. Variations of this architecture include integrating automatic identification system (AIS) \cite{aiAISDataAided2022,miliosAutomaticFusionSatellite2019,dengSARSHIPNETSARSHIPDETECTION2022,beretaVesselDetectionUsing2020} data to correct for false detections or to aid in dataset quality, while others use variations of the CFAR method, such as a truncated-clutter-statistics-based joint CFAR \cite{aiMultiScaleRotationInvariantHaarLike2019} algorithm, to enhance pre-screening success in a multi-target analysis environment.

The field of computer vision has advanced significantly through methods initially built for medical imaging. We see these techniques also applied to detect ships in satellite imagery. Techniques implementing radon transforms, which are widely used in tomography, map individual pixels to polar coordinates. \cite{corbaneCompleteProcessingChain2010} This transform maps individual pixels to polar coordinates to obtain orientation and spacing. Most importantly, it is reversible. This allows the image to be reconstructed after obtaining spatial information. Similarly, a U-Net architecture, which classifies all pixels in an image, is used in a number of these techniques and is commonly used in biomedical imaging. \cite{samarEnhancingPerformanceIR2021,gubelloImageProcessingTraditional2022} Rather than classifying an entire image into a single category, this architecture enables image segmentation to partition the imagery. This is important to perform robust ship detection, as there can be multiple potential targets within a single image.

More recently, techniques utilizing vision transformers \cite{dosovitskiyImageWorth16x162021} have been used to perform ship detection \cite{guptaXTNestedTokenization2024}. Vision transformers utilize a simple mechanism, attention \cite{bahdanauNeuralMachineTranslation2014}, stacked repeatedly to learn effective representations from images. Particularly, they scale well as the models get larger, demonstrating state-of-the-art performance in computer vision today and the basis of modern vision foundation models.

Despite the prevalence of vision transformers and their capability for superior performance, the simplicity of CNN-based methods and the wide availability of easy tools to train CNN-based object detection models lead to the dominance of CNNs in modern ship detection methods. It is our view that Chinese naval planners are likely using these capabilities to characterize U.S. naval assets.

\subsection{Potential \textit{Zhousidun} Dataset Use-Cases and Value}
The \textit{Zhousidun} dataset is mainly composed of images of \textit{Arleigh Burke}-class destroyers, with some other destroyers and frigates belonging to other navies. The \textit{Arleigh Burke}-class make up the vast majority of the database, likely due to their strong presence in the Indo-Pacific and their large numbers in the U.S. Navy.

\textit{Arleigh Burke} destroyers have a number of different operational tasks within the U.S. Navy that are crucial to U.S. military operations in the Indo-Pacific.\footnote{The Arleigh-Burke class of destroyers is the main destroyer for the U.S. Navy. The only other destroyers in the Navy are the 3 Zumwalt destroyers, of which only 2 are commissioned.} In particular, the ships are heavily relied upon for conducting freedom of navigation operations (FONOPs) \cite{commanderu.s.7thfleetpublicaffairsNavyDestroyerConducts2024}, defending U.S. assets, particularly carrier groups, with missile and air defense, and conducting anti-submarine warfare (ASW). \textit{Arleigh Burke} destroyers, with their Aegis missile defense systems, could also serve to provide theater air defense to provide coverage for combat operations. This missile defense capability is particularly important in the face of growing PRC conventional ballistic missile capabilities, including Anti-Ship Ballistic Missiles (ASBMs) that can hold U.S. carriers and carrier groups at risk.\footnote{``The number of available Anti-ship Ballistic Missiles (ASBMs) has likely already broadened the PLARF's anti-ship mission from what has been thought of as a 'carrier-killer' role to a broader and more generic 'ship-killer' mission.'' \cite{shugartDeterringPowerfulEnemy2024}}

\textit{Arleigh Burke} destroyers are thus an immediate threat to adversary operations and likely to be targeted in conflict, both because of their own operations (like ASW) but due to their role as defensive pickets for fleet and theater operations. Even outside of conflict, as the most numerous class of U.S. warships, their breadth of operation, particularly in the Indo-Pacific, means that they are often coming into contact with ships from competitor navies,\footnote{Including some close calls with the PLAN: \href{https://www.cbsnews.com/news/chinese-warship-u-s-missile-destroyer-taiwan-strait-close-call-uss-chung-hoon/}{Chinese warship comes within 150 yards of U.S. missile destroyer in Taiwan Strait - CBS News}; \href{https://apnews.com/article/us-china-destroyer-ship-paracel-south-china-sea-a146fa95de728afb0b3a9d8a49dceeb8}{US denies Chinese claim it drove away American destroyer - AP News}} conducting exercises with allies \cite{u.s.7thfleetpublicaffairsJapanConductBilateral2024}, and part of any changes or evolution in American maritime strategy.

Specifically, \textit{Zhousidun} appears to focus on key Aegis components of \textit{Arleigh Burke} ships with their bounding boxes. Models trained to detect the Aegis combat system can be used to both directly target the combat system itself\footnote{The Aegis combat system is a powerful warfare capability but is not the only component that makes an \textit{Arleigh Burke}-class destroyer useful. The use of electronic warfare, precision munitions, or small drones to destroy \textit{Aegis} components specifically could be a less escalatory act than targeting a vessel directly. Models that can detect these components specifically would aid in the use of such weapons.} or to improve the detection performance of destroyers at large. Training a model to detect and discriminate between different categories of naval vessels can be tricky and cumbersome as a model can confuse destroyers for other types of vessels. If detecting destroyers with high recall is of priority, then it may be prudent to instead detect a unique component of destroyers specifically. This would reduce recall on non-destroyer vessels to zero while potentially improving precision for only destroyers. This approach is already in use by multiple defense contractors in the United States for tasks such as human activity recognition.\footnote{\href{https://oksi.ai/defense/\#computer}{OKSI AI} detects not only people, but also components on that person to identify their behavior.}

\section{Why This Matters: Adaptive Net Assessment and the Open-Source}

In addition to the specifics of the \textit{Zhoushidun} dataset leak, this episode underscores the potential incorporation of AI into net assessment. Machine learning capabilities are poised to become a core component of many military programs in the near future—particularly as it relates to intelligence, surveillance, and reconnaissance. Understanding how these capabilities work and how well they work represents a central part of the net assessment capabilities for any nation. 

With the growing significance of quantitative capabilities, net assessment must integrate new quantitative methods for model evaluation. We can reconstruct theoretical workflows for analysis by utilizing knowledge of existing data and deployment workflows. While qualitative insights into bureaucratic politics and command control remain crucial, incorporating quantitative techniques enhances our understanding of how technical constraints impact the military balance.

We recommend adopting and scaling the simulation and mirroring approach to the assessment of machine learning models exemplified here. State-of-the-art models are available in the open source and can be obtained and executed cheaply as a white box. Input data to train these models can be created with high fidelity using simulation and modeling tools or collected from the open source, as in this case. Critically, the entire pipeline can be modified at will to understand how varying inputs affect the veracity and reliability of the models being assessed. OSINT analysts can be deployed at many agencies, governmental or non-governmental, to unearth datasets that might be used to train machine learning models to bootstrap the net assessment of a given capability.

Open-source intelligence can provide important insight into how public academic techniques are evolving, as well as any public techniques from the private sector. This is a critical insight as developments in the academic and commercial AI sector are a harbinger for military AI capabilities.
Here, we demonstrate that an analysis of the \textit{Zhoushidun} dataset as a quantitative assessment capability represents a potentially useful predictor of China’s machine-learning arsenal. We hope that replication of this method using similar candidate datasets might present new opportunities for the future net assessment of AI models and use cases. Repeated use of such a methodology on new datasets will provide tighter bounds and more accurate assessment of competitor AI capabilities.

\section*{Acknowledgements}
We would like to acknowledge the support of the Founder’s Pledge Fund for their support of the Berkeley Risk and Security Laboratory and this work. We thank Trevor Darrell, Suzanne Petryk, and other colleagues for their comments and feedback on this publication.

\clearpage

\begin{appendices}
\onecolumn
\section{Appendix A: Model Configuration}
\label{app:data-model}

The \textit{Zhousidun} dataset is split into train and validation sets of 557 and 51 images respectively. For the purposes of this work, the validation set is used as an external test set and is not used for training or hyperparameter tuning.

\begin{figure}
    \centering
    \includegraphics{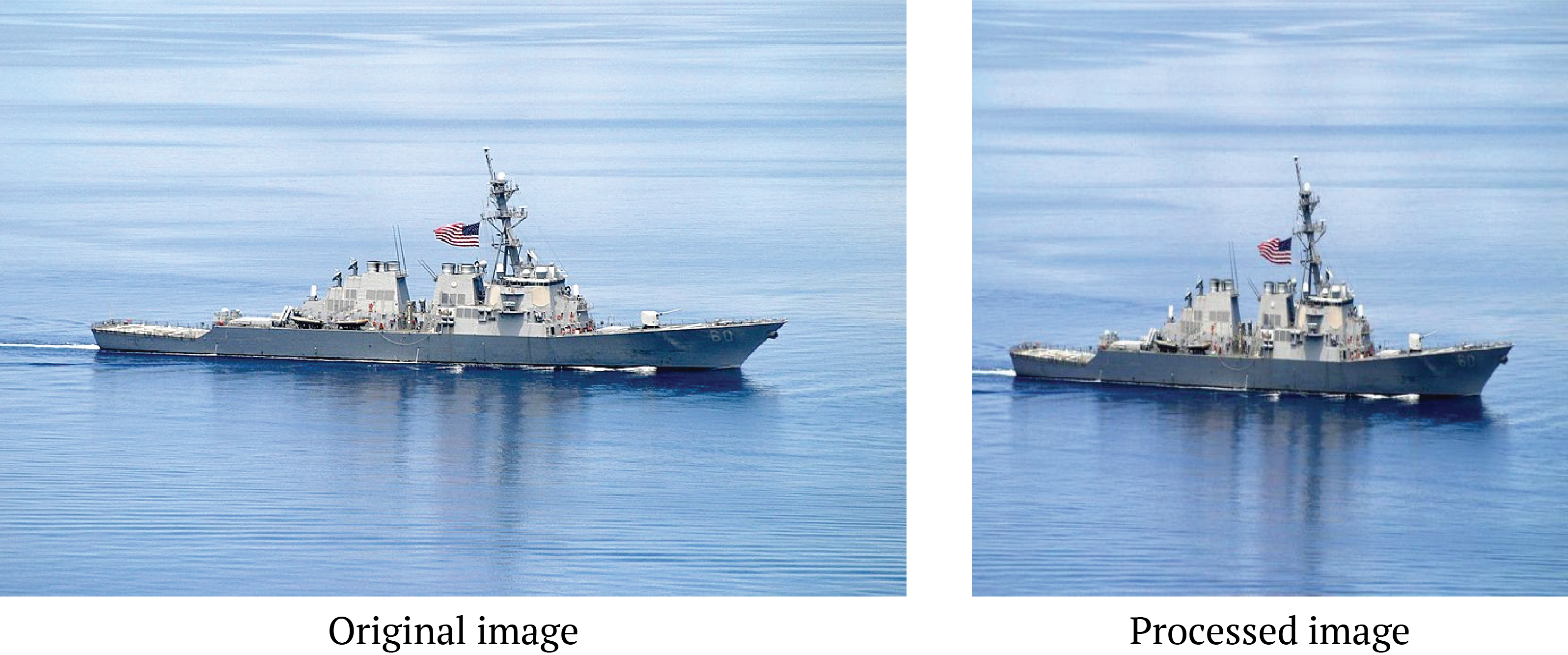}
    \caption{(left) An image of the USS Paul Hamilton (DDG-60) as originally released by the U.S. Navy at 2,274 × 1,494 pixel resolution. (right) The same image as part of the Zhousidun dataset processed to 640 x 640 pixel resolution. Notice that the shape of the ship and its components are drastically altered.}
    \label{fig:hi-los-res}
\end{figure}

The original images are all of varying resolutions. However, a limitation of the YOLO-family of architectures is that input images must all be of the same size. Therefore, each image is resized to be a square 640 x 640px resolution by the Roboflow platform, as visualized in Figure \ref{fig:hi-los-res}. This may have negative consequences on the performance of the model as the aspect ratio of the Aegis components is sometimes drastically changed, but is otherwise a common practice when using CNN-based object detectors.

The YOLOv8l model used in this work is trained for 200 epochs using the AdamW optimizer. The Ultralytics implementation utilized a heuristic to set an optimal learning rate for the optimizer, resulting in a learning rate of $\alpha$=7.14e-4 and momentum $\beta$=0.9. We utilize parameter group weight decay with parameter groups, set automatically by the Ultralytics library:

\begin{itemize}
  \item 97 weight decay=0.0
  \item 104 weight decay=0.0015625
  \item 103 bias decay=0.0
\end{itemize}

\begin{figure*}[h]
    \centering
    \includegraphics[width=\textwidth]{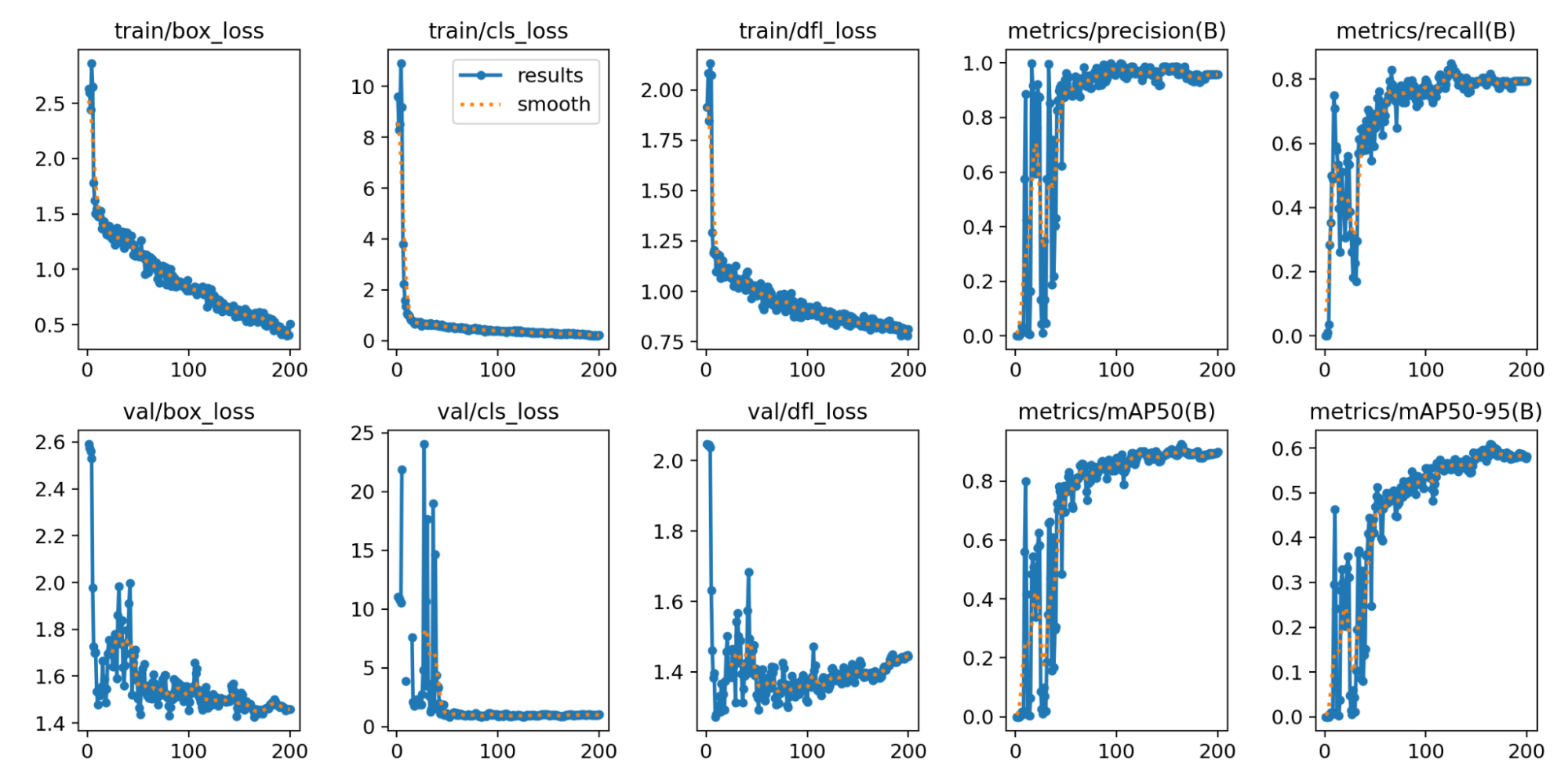}
    \caption{Training loss and metric curves for the reported YOLOv8l model.}
    \label{fig:app-loss}
\end{figure*}

The training is accelerated with 10 NVIDIA P100s with a batch size of 200. The gain in precision, recall, and mAP plateaus approximately at epoch 150. We continue the training for the full 200 epochs for completeness but report the highest mAP achieved during the training run. Loss curves for the training run are provided in Figure \ref{fig:app-loss}.

\clearpage

\twocolumn
\section{Appendix B: Ships Identified in \textit{Zhousidun}}
\label{app:vessels}

We created a list of all identifiable ships in the \textit{Zhousidun} dataset using the hull classification numbers visible in the images. Many images had unreadable hull classification numbers either due to poor image quality or capture angle. Therefore, this list may not be complete. 

\subsection*{United States of America}
\begin{itemize}
  \item USS Arleigh Burke (DDG-51)
  \item USS Barry (DDG-52)
  \item USS John Paul Jones (DDG-53)
  \item USS Curtis Wilbur (DDG-54)
  \item USS Stout (DDG-55)
  \item USS John S. McCain (DDG-56)
  \item USS Mitscher (DDG-57)
  \item USS Paul Hamilton (DDG-60)
  \item USS Ramage (DDG-61)
  \item USS Fitzgerald (DDG-62)
  \item USS Carney (DDG-64)
  \item USS Benfold (DDG-65)
  \item USS Gonzalez (DDG-66)
  \item USS Cole (DDG-67)
  \item USS The Sullivans (DDG-68)
  \item USS Milius (DDG-69)
  \item USS Hopper (DDG-70)
  \item USS Magan (DDG-72)
  \item USS Donald Cook (DDG-75)
  \item USS Porter (DDG-78)
  \item USS Roosevelt (DDG-80)
  \item USS Winston S. Churchill (DDG-81)
  \item USS Howard (DDG-83)
  \item USS Bulkeley (DDG-84)
  \item USS McCampbell (DDG-85)
  \item USS Shoup (DDG-86)
  \item USS Mason (DDG-87)
  \item USS Preble (DDG-88)
  \item USS Mustin (DDG-89)
  \item USS Chafee (DDG-90)
  \item USS Pinckney (DDG-91)
  \item USS Momsen (DDG-92)
  \item USS Chung-Hoon (DDG-93)
  \item USS Bainbridge (DDG-96)
  \item USS Farragut (DDG-99)
  \item USS Kidd (DDG-100)
  \item USS Gridley (DDG-101)
  \item USS Truxtun (DDG-103)
  \item USS Sterett (DDG-104)
  \item USS Dewey (DDG-105)
  \item USS Stockdale (DDG-106)
  \item USS Gravely (DDG-107)
  \item USS Jason Dunham (DDG-109)
  \item USS William P. Lawrence (DDG-110)
  \item USS Spruance (DDG-111)
  \item USS Michael Murphy (DDG-112)
  \item USS Rafael Peralta (DDG-115)
  \item USS Paul Ignatius (DDG-117)
\end{itemize}
\subsection*{Japan}
\begin{itemize}
  \item JS Kongō (DDG-173)
  \item JS Kirishima (DDG-174)
  \item JS Myōkō (DDG-175)
  \item JS Atago (DDG-177)
  \item JS Chōkai (DDG-176)
  \item JS Ashigara (DDG-178)
\end{itemize}
\subsection*{Republic of Korea}
\begin{itemize}
  \item ROKS Munmu the Great (DDH-976)
  \item ROKS Wang Geon (DDH-978)
  \item ROKS Sejong the Great (DDG-991)
  \item ROKS Yulgok Yi I (DDG-992)
\end{itemize}
\subsection*{Norway}
There was only one image of a Norwegian naval vessel (frigate) in the dataset.

\begin{itemize}
  \item HNoMS Thor Heyerdahl (F314)
\end{itemize}
\subsection*{Spain}
There was only one image of a Spanish naval vessel (frigate) in the dataset.

\begin{itemize}
  \item Blas de Lezo (F-103)
\end{itemize}
\subsection*{Australia}
There was only one image of an Australian naval vessel (destroyer) in the dataset.

\begin{itemize}
  \item HMAS Hobart (DDG-39)
\end{itemize}
\subsection*{People’s Republic of China}
There were only a total of three images of Chinese naval vessels in the dataset.

\begin{itemize}
  \item Jinan (152)
  \item Lanzhou (170)
  \item Kunming (172)
\end{itemize}

\clearpage

\section{Appendix C: Papers Reviewed}
\label{app:papers}

\begin{flushleft}
A Comparison of Fixed Threshold CFAR and CNN Ship Detection Methods for S-band NovaSAR Images \cite{carmanComparisonFixedThreshold2020}
\begin{itemize}
  \item Researcher Affiliations: Defence Science and Technology Laboratory.
  \item Country of Institutions: United Kingdom.
\end{itemize}

A Complete Processing Chain For Ship Detection Using Optical Satellite Imagery \cite{corbaneCompleteProcessingChain2010}
\begin{itemize}
  \item Researcher Affiliations: ESPACE Unit, Institut de Recherche pour le Développement; Laboratoire d’Informatique Gaspard-Monge, Université Paris-Est; IUP Génie Physiologique et Informatique Faculté des Sciences Fondamentales et Appliquées, Université de Poitiers.
  \item Country of Institutions: France.
\end{itemize}

A Deep-Learning-Based Lightweight Model for Ship Localizations in SAR Images \cite{bhattacharjeeDeepLearningBasedLightweightModel2023}
\begin{itemize}
  \item Researcher Affiliations: Department of Ocean Engineering, Indian Institute of Technology Madras; Department of Computer Science, Indian Institute of Technology Madras.
  \item Country of Institutions: India.
\end{itemize}

A Novel Full-Polarization SAR Images Ship Detector Based on the Scattering Mechanisms and the Wave Polarization Anisotropy \cite{zhangNovelFullPolarizationSAR2021}
\begin{itemize}
  \item Researcher Affiliations: Faculty of Geosciences and Environmental Engineering, Southwest Jiaotong University, Chengdu; the National University of Defense Technology, Changsha; Hunan University; Naval Aviation University of China.
  \item Country of Institutions: China.
\end{itemize}

A Novel Hierarchical Method of Ship Detection from Spaceborne Optical Image Based on Shape and Texture Features \cite{zhuNovelHierarchicalMethod2010}
\begin{itemize}
  \item Researcher Affiliations: ATR National Laboratory, National University of Defense Technology, Changsha; ATR National Laboratory, School of Electrical Science and Engineering, National University of Defense Technology, Changsha.
  \item Country of Institutions: China.
\end{itemize}

A SAR Dataset of Ship Detection for Deep Learning under Complex Backgrounds \cite{wangSARDatasetShip2019}
\begin{itemize}
  \item Researcher Affiliations: Key Laboratory of Digital Earth Science, Institute of Remote Sensing and Digital Earth, Chinese Academy of Sciences; University of Chinese Academy of Sciences.
  \item Country of Institutions: China.
\end{itemize}

ABOShips – An Inshore and Offshore Maritime Vessel Detection Dataset with Precise Annotations \cite{iancuABOShipsInshoreOffshore2021}
\begin{itemize}
  \item Researcher Affiliations: Faculty of Science and Engineering, Åbo Akademi University.
  \item Country of Institutions: Finland.
\end{itemize}

AIS Data Aided Rayleigh CFAR Ship Detection Algorithm of Multiple-Target Environment in SAR Images \cite{aiAISDataAided2022}
\begin{itemize}
  \item Researcher Affiliations: Hefei University of Technology; China Electronics Technology Corporation; Hebei University of Technology; Tianjin Institute of Advanced Technology; Xidian University.
  \item Country of Institutions: China.
\end{itemize}

AMANet: Advancing SAR Ship Detection With Adaptive Multi-Hierarchical Attention Network \cite{maAMANetAdvancingSAR2023}
\begin{itemize}
  \item Researcher Affiliations: Shijiazhuang Campus, Army Engineering University of the People’s Liberation Army; School of Automation, Northwestern Polytechnical University.
  \item Country of Institutions: China.
\end{itemize}

Arbitrary-Oriented Ship Detection through Center-Head Point Extraction \cite{zhangArbitraryOrientedShipDetection2021}
\begin{itemize}
  \item Researcher Affiliations: College of Electronic Science and Technology, National University of Defense Technology, Changsha.
  \item Country of Institutions: China.
\end{itemize}

Automatic Fusion of Satellite Imagery and AIS Data for Vessel Detection \cite{miliosAutomaticFusionSatellite2019}
\begin{itemize}
  \item Researcher Affiliations: Institute of Big Data Analytics, Dalhousie University; MarineTraffic; University of the Aegean; Institute of Computer Science, Polish Academy of Sciences.
  \item Country of Institutions: Canada, United Kingdom, Greece, Poland.
\end{itemize}

Automatic Offshore Infrastructure Extractions Using Synthetic Aperture Radar \& Google Earth Engine \cite{wongAutomatingOffshoreInfrastructure2019}
\begin{itemize}
  \item Researcher Affiliations: Marine Geospatial Ecology Lab, Nicholas School of the Environment, Duke University; SkyTruth.
  \item Country of Institutions: United States.
\end{itemize}

Compressed-Domain Ship Detection on Spaceborne Optical Image Using Deep Neural Network and Extreme Learning Machine \cite{tangCompresseddomainShipDetection2015}
\begin{itemize}
  \item Researcher Affiliations: School of Information and Electronics, Beijing Institute of Technology; School of Electrical and Electronic Engineering, Nanyang Technological University.
  \item Country of Institutions: China, Singapore.
\end{itemize}

Counting from Sky: A Large-scale Dataset for Remote Sensing Object Counting and A Benchmark Method \cite{gaoCountingSkyLargescale2020}
\begin{itemize}
  \item Researcher Affiliations: State Key Laboratory of Virtual Reality Technology and Systems, Beihang University; Hangzhou Innovation Institute, Beihang University.
  \item Country of Institutions: China.
\end{itemize}

CVGG-Net: Ship Recognition for SAR Images Based on Complex-Valued Convolutional Neural Network \cite{zhaoCVGGNetShipRecognition}
\begin{itemize}
  \item Researcher Affiliations: School of Information and Communication Engineering, Hainan University; Suzhou Key Laboratory of Microwave Imaging, Processing and Application Technology;  Suzhou Aerospace Information Research Institute; Aerospace Information Research Institute, Chinese Academy of Sciences; Soochow University.
  \item Country of Institutions: China.
\end{itemize}

Deep Convolutional Neural Network based Ship Images Classification \cite{mishraDeepConvolutionalNeural2021}
\begin{itemize}
  \item Researcher Affiliations: Weapons and Electronics Systems Engineering Establishment, New Delhi.
  \item Country of Institutions: India.
\end{itemize}

Deep Learning for Autonomous Ship-Oriented Small Ship Detection \cite{chenDeepLearningAutonomous2020}
\begin{itemize}
  \item Researcher Affiliations: Intelligent Transportation Systems Research Center, Wuhan University of Technology; School of Computer Science and Technology, Wuhan University of Technology; School of Management, Wuhan University of Technology; School of Engineering, Department of Mechanical Engineering, Marine Technology, Aalto University.
  \item Country of Institutions: China, Finland.
\end{itemize}

Deep Learning-Based Automatic Detection of Ships: An Experimental Study Using Satellite Images \cite{patelDeepLearningBasedAutomatic2022}
\begin{itemize}
  \item Researcher Affiliations: Department of Computer Science \& Engineering, Devang Patel Institute of Advance Technology and Research (DEPSTAR), CHARUSAT Campus, Charotar University of Science and Technology; U \& P U. Patel Department of Computer Engineering, Chandubhai S Patel Institute of Technology, CHARUSAT Campus, Charotar University of Science and Technology; Institute of Applied Sciences and Intelligent Systems, National Research Council of Italy.
  \item Country of Institutions: India, Italy.
\end{itemize}

Deep Learning-Based Ship Detection in Remote Sensing Imagery Using TensorFlow \cite{apoorvaDeepLearningBasedShip2020}
\begin{itemize}
  \item Researcher Affiliations: Computer Science and Engineering, Parala Maharaja Engineering College; Department of Basic Science, Parala Maharaja Engineering College.
  \item Country of Institutions: India.
\end{itemize}

Design and Evaluation of a Artificial Intelligence Based Vessel Detection System in Pol-SAR Images \cite{benito-ortizDesignEvaluationArtificial2018}
\begin{itemize}
  \item Researcher Affiliations: Universidad de Alcalá; Signal Theory and Communications Department Polytechnic School, Universidad de Alcalá.
  \item Country of Institutions: Spain.
\end{itemize}

DOTA: A Large-Scale Dataset for Object Detection in Aerial Images \cite{xiaDOTALargescaleDataset2018}
\begin{itemize}
  \item Researcher Affiliations: Wuhan University.
  \item Country of Institutions: China.
\end{itemize}

Enhancing Performance of IR Ship Detection with Baseline AI Models Over a New Benchmark \cite{samarEnhancingPerformanceIR2021}
\begin{itemize}
  \item Researcher Affiliations: Department of Computer Science, National University of Computer and Emerging Sciences; Department of Control and Signal Processing, Center of Excellence in Science and Applied Technologies.
  \item Country of Institutions: Pakistan.
\end{itemize}

Hierarchical and Robust Convolutional Neural Network for Very High-Resolution Remote Sensing Object Detection \cite{zhangHierarchicalRobustConvolutional2019}
\begin{itemize}
  \item Researcher Affiliations: School of Electronic, Electrical and Communication Engineering, University of Chinese Academy of Sciences; Key Laboratory of Spectral Imaging Technology, Xi’an Institute of Optics and Precision Mechanics, Chinese Academy of Sciences.
  \item Country of Institutions: China.
\end{itemize}

Image Processing with Traditional and Artificial Intelligence Techniques for Ship Detection \cite{gubelloImageProcessingTraditional2022}
\begin{itemize}
  \item Researcher Affiliations: Politecnico di Milano.
  \item Country of Institutions: Italy.
\end{itemize}

Large-Scale Automatic Vessel Monitoring Based on Dual-Polarization Sentinel-1 and AIS Data \cite{pelichLargescaleAutomaticVessel2019}
\begin{itemize}
  \item Researcher Affiliations: Luxembourg Institute of Science and Technology, Environmental Research and Innovation Department; LuxSpace Sàrl.
  \item Country of Institutions: Luxembourg.
\end{itemize}

LS-SSDD-v1.0: A Deep Learning Dataset Dedicated to Small Ship Detection from Large-Scale Sentinel-1 SAR Images \cite{zhangLSSSDDv1DeepLearning2020}
\begin{itemize}
  \item Researcher Affiliations: School of Information and Communication Engineering, University of Electronic Science and Technology of China; Aerospace Information Research Institute, Chinese Academy of Sciences; Department of Electronic and Information Engineering, Naval Aeronautical University; School of Electronic Information and Electrical Engineering, Shanghai Jiao Tong University.
  \item Country of Institutions: China.
\end{itemize}

Multi-Scale Rotation-Invariant Haar-Like Feature Integrated CNN-Based Ship Detection Algorithm of Multiple-Target Environment in SAR Imagery \cite{aiMultiScaleRotationInvariantHaarLike2019}
\begin{itemize}
  \item Researcher Affiliations: Key Laboratory of Knowledge Engineering With Big Data, Hefei University of Technology; Ministry of Education, Hefei; School of Computer Science and Information Engineering, Hefei University of Technology; School of Automation, Central South University; Department of Electrical and Computer Engineering, Mississippi State University.
  \item Country of Institutions: China, United States.
\end{itemize}

NovaSAR and SSTL S1-4: SAR and EO Data Fusion \cite{couchman-crookNovaSARSSTLS142020}
\begin{itemize}
  \item Researcher Affiliations: Defence Science and Technology Laboratory.
  \item Country of Institutions: United Kingdom.
\end{itemize}

NWPU-MOC: A Benchmark for Fine-grained Multi-category Object Counting in Aerial Images \cite{gaoNWPUMOCBenchmarkFinegrained2021}
\begin{itemize}
  \item Researcher Affiliations: School of Artificial Intelligence, Optics and Electronics, Northwestern Polytechnical University; Key Laboratory of Intelligent Interaction and Applications, Ministry of Industry and Information Technology, Xi’an.
  \item Country of Institutions: China.
\end{itemize}

Object Detection in Aerial Images: A Large-Scale Benchmark and Challenges \cite{dingObjectDetectionAerial2021}
\begin{itemize}
  \item Researcher Affiliations: State Key Laboratory of Information Engineering in Surveying, Mapping and Remote Sensing, Wuhan University; National Engineering Research Center for Multimedia Software, School of Computer Science and Institute of Artificial Intelligence, Wuhan University; School of Electronic Information, Huazhong University; Faculty of Geo-Information, Wuhan University; Department of Computer Science, Cornell University and Cornell Tech; Department of Computer Science, University of Rochester; Remote Sensing Technology Institute, German Aerospace Center; University of POLITEHNICA of Bucharest; DAIS, Ca’ Foscari University of Venice.
  \item Country of Institutions: China, United States, Germany, Romania, Italy.
\end{itemize}

Object-based image analysis approach for vessel detection on optical and radar images \cite{aielloObjectbasedImageAnalysis2019}
\begin{itemize}
  \item Researcher Affiliations: Politecnico di Milano.
  \item Country of Institutions: Italy.
\end{itemize}

Objects as Points \cite{zhouObjectsPoints2019}
\begin{itemize}
  \item Researcher Affiliations: University of Texas, Austin; University of California, Berkeley.
  \item Country of Institutions: United States.
\end{itemize}

Research on Mosaic Image Data Enhancement for Overlapping Ship Targets \cite{zengResearchMosaicImage2021}
\begin{itemize}
  \item Researcher Affiliations: Marine Engineering Institute, Jimei University; Shanghai Maritime University; School of Marine Engineering, Jimei University.
  \item Country of Institutions: China.
\end{itemize}

Rotated Region Based CNN for Ship Detection \cite{liuRotatedRegionBased2017}
\begin{itemize}
  \item Researcher Affiliations: University of Chinese Academy of Sciences; Institute of Automation, Chinese Academy of Sciences.
  \item Country of Institutions: China.
\end{itemize}

SAR Ship Target Recognition Via Multi-Scale Feature Attention and Adaptive-Weighted Classifier \cite{wangSARShipTarget2023}
\begin{itemize}
  \item Researcher Affiliations: Department of Electrical Engineering, University of Electronic Science and Technology of China.
  \item Country of Institutions: China.
\end{itemize}

SAR-Shipnet: SAR-Ship Detection Neural Network Via Bidirectional Coordinate Attention and Multi-Resolution Feature Fusion \cite{dengSARSHIPNETSARSHIPDETECTION2022}
\begin{itemize}
  \item Researcher Affiliations: School of Mechanical Engineering and Automation, Harbin Institute of Technology; Kyung Hee University; Department of Computer Science, Tsinghua University; College of Computer Science and Technology, Nanjing University of Aeronautics and Astronautics; School of Mechanical and Electrical Engineering, Nanjing University of Aeronautics and Astronautics.
  \item Country of Institutions: China, Republic of Korea.
\end{itemize}

Satellite Image Recognition for Smart Ships Using a Convolutional Neural Networks Algorithm \cite{xiaoSatelliteImageRecognition2019}
\begin{itemize}
  \item Researcher Affiliations: State Street Global Markets, State Street Corporation; Data Science Institute, Saint Peter’s University; Big Data and AI Lab, IntelligentRabbit LLC.
  \item Country of Institutions: United States.
\end{itemize}

SeaShips: A Large-Scale Precisely Annotated Dataset for Ship Detection \cite{shaoSeaShipsLargescalePrecisely2018}
\begin{itemize}
  \item Researcher Affiliations: State Key Laboratory for Information Engineering in Surveying, Mapping and Remote Sensing, Wuhan University; National Engineering Research Center for Multimedia Software, Wuhan University; Computer Science and Engineering, University of California, Merced.
  \item Country of Institutions: China, United States.
\end{itemize}

Ship Detection Based on YOLOv2 for SAR Imagery \cite{changShipDetectionBased2019}
\begin{itemize}
  \item Researcher Affiliations: Department of Electrical Engineering, National Taipei University of Technology; Department of Communications and Guidance Engineering, National Taiwan Ocean University.
  \item Country of Institutions: Taiwan.
\end{itemize}

Ship Detection for High-Resolution SAR Images Based on Feature Analysis \cite{wangShipDetectionHighResolution2014}
\begin{itemize}
  \item Researcher Affiliations: Chinese Academy of Sciences.
  \item Country of Institutions: China.
\end{itemize}

Ship Detection from Optical Satellite Images Based on Sea Surface Analysis \cite{yangShipDetectionOptical2014}
\begin{itemize}
  \item Researcher Affiliations: State Key Laboratory of Virtual Reality Technology and System, Beihang University.
  \item Country of Institutions: China.
\end{itemize}

Ship Detection in SAR Images Based on an Improved Faster R-CNN \cite{liShipDetectionSAR2017}
\begin{itemize}
  \item Researcher Affiliations: Department of Electronic and Information Engineering, Naval Aeronautical and Astronautical University.
  \item Country of Institutions: China.
\end{itemize}

Ship Detection in SAR Images with Human-in-the-Loop \cite{jiaShipDetectionSAR}
\begin{itemize}
  \item Researcher Affiliations: Key Laboratory of Information Science of Electromagnetic Waves, Fudan University.
  \item Country of Institutions: China.
\end{itemize}

Ship Detection with Spectral Analysis of Synthetic Aperture Radar: A Comparison of New and Well-Known Algorithms \cite{marinoShipDetectionSpectral2015}
\begin{itemize}
  \item Researcher Affiliations: Institute of Environmental Engineering, ETH Zurich; The Open University, Department of Engineering and Innovation; Microwaves and Radar Institute, German Aerospace Center; Korea Institute of Ocean Science and Technology, Korea Ocean Satellite Center.
  \item Country of Institutions: Switzerland, United Kingdom, Germany, Republic of Korea.
\end{itemize}

SISP: A Benchmark Dataset for Fine-Grained Ship Instance Segmentation in Panchromatic Satellite Images \cite{fengSISPBenchmarkDataset2024}
\begin{itemize}
  \item Researcher Affiliations: State Key Laboratory of Space-Ground Integrated Information Technology, CAST; Group of Intelligent Signal Processing, College of Computer Science and Technology, Harbin Engineering University; College of Electrical and Mechanical Engineering, Dalian Minzu University; School of Geography and Ocean Science, Nanjing University.
  \item Country of Institutions: China.
\end{itemize}

The digital frontiers of fisheries governance: fish attraction devices, drones and satellites \cite{toonenDigitalFrontiersFisheries2020}
\begin{itemize}
  \item Researcher Affiliations: Environmental Policy Group, Wageningen University.
  \item Country of Institutions: Netherlands.
\end{itemize}

Towards Arbitrary-Oriented Ship Detection With Rotated Region Proposal and Discrimination Networks \cite{zhangArbitraryorientedShipDetection2018}
\begin{itemize}
  \item Researcher Affiliations: Shanghai Key Laboratory of Intelligent Sensing and Recognition, Shanghai Jiao Tong University.
  \item Country of Institutions: China.
\end{itemize}

Towards an Explainable Artificial Intelligence Approach for Ships Detection from Satellite Imagery \cite{ieracitanoExplainableArtificialIntelligence2023}
\begin{itemize}
  \item Researcher Affiliations: DICEAM, University Mediterranea of Reggio Calabria.
  \item Country of Institutions: Italy.
\end{itemize}

Vessel Detection using Image Processing and Neural Networks \cite{beretaVesselDetectionUsing2020}
\begin{itemize}
  \item Researcher Affiliations: MarineTraffic; NATO-STO-CMRE; University of the Aegean.
  \item Country of Institutions: Greece, Italy.
\end{itemize}

xView: Objects in Context in Overhead Imagery \cite{lamXViewObjectsContext2018}
\begin{itemize}
  \item Researcher Affiliations: Defense Innovation Unit Experimental; DigitalGlobe; National Geospatial-Intelligence Agency.
  \item Country of Institutions: United States.
\end{itemize}

xView3-SAR: Detecting Dark Fishing Activity Using Synthetic Aperture Radar Imagery \cite{paoloXView3SARDetectingDark2022}
\begin{itemize}
  \item Researcher Affiliations: Global Fishing Watch, Cambrio, Defense Innovation Unit, University of California, Berkeley.
  \item Country of Institutions: United States.
\end{itemize}

OpenSARShip: A Dataset Dedicated to Sentinel-1 Ship Interpretation \cite{liOpenSARShipLargevolumeDataset2017}
\begin{itemize}
  \item Researcher Affiliations: Shanghai Key Laboratory of Intelligent Sensing and Recognition, Shanghai Jiao Tong University.
  \item Country of Institutions: China.
\end{itemize}

OpenSARShip 2.0: A Large-Volume Dataset for Deeper Interpretation of Ship Targets in Sentinel-1 Imagery \cite{liOpenSARShipLargevolumeDataset2017}
\begin{itemize}
  \item Researcher Affiliations: Shanghai Key Laboratory of Intelligent Sensing and Recognition, Shanghai Jiao Tong University.
  \item Country of Institutions: China.
\end{itemize}

AIR-SARShip-1.0: High-resolution SAR Ship Detection Dataset \cite{xianAIRSARShip1HighresolutionSAR2019}
\begin{itemize}
  \item Researcher Affiliations: Aerospace Information Research Institute, Chinese Academy of Sciences; University of Chinese Academy of Sciences; Key Laboratory of Network Information System Technology.
  \item Country of Institutions: China.
\end{itemize}

SAR-AIRcraft-1.0: High-resolution SAR aircraft detection and recognition dataset \cite{wangzhiruiSARAIRcraft1HighresolutionSAR2023}
\begin{itemize}
  \item Researcher Affiliations: Aerospace Information Research Institute, Chinese Academy of Sciences; University of Chinese Academy of Sciences; School of Electronic, Electrical and Communication Engineering, University of Chinese Academy of Sciences; Key Laboratory of Network Information System Technology (NIST), Chinese Academy of Sciences.
  \item Country of Institutions: China.
\end{itemize}

HRSID: A High-Resolution SAR Images Dataset for Ship Detection and Instance Segmentation \cite{weiHRSIDHighResolutionSAR2020}
\begin{itemize}
  \item Researcher Affiliations: School of Information and Communication Engineering, University of Electronic Science and Technology of China.
  \item Country of Institutions: China.
\end{itemize}

FUSAR-Ship: building a high-resolution SAR-AIS matchup dataset of Gaofen-3 for ship detection and recognition \cite{houFUSARShipBuildingHighresolution2020}
\begin{itemize}
  \item Researcher Affiliations: Key Laboratory for Information Science of Electromagnetic Waves, Fudan University; Shanghai Center for Gaofen Data and Applications.
  \item Country of Institutions: China.
\end{itemize}

SAR Ship Detection Dataset (SSDD): Official Release and Comprehensive Data Analysis \cite{zhangSARShipDetection2021}
\begin{itemize}
  \item Researcher Affiliations: School of Information and Communication Engineering, University of Electronic Science and Technology of China; Department of Electronic and Information Engineering, Naval Aeronautical University; Department of Electrical and Electronic Engineering, University of Hong Kong; Dahua Technology; State Key Laboratory of Information Engineering in Surveying, Mapping, and Remote Sensing, Wuhan University; Aerospace Information Research Institute, Chinese Academy of Sciences; College of Information Science and Technology, Dalian Maritime University; School of Electronic Information and Electrical Engineering, Shanghai Jiao Tong University.
  \item Country of Institutions: China.
\end{itemize}

A Dual-Polarimetric SAR Ship Detection Dataset and a Memory-Augmented Autoencoder-Based Detection Method \cite{huDualPolarimetricSARShip2021}
\begin{itemize}
  \item Researcher Affiliations: Aerospace Research Institute, Chinese Academy of Sciences; Key Laboratory of Technology in Geo-Spatial Information Processing and Application System, Chinese Academy of Sciences; School of Electronic, Electrical and Communication Engineering, University of Chinese Academy of Sciences.
  \item Country of Institutions: China.
\end{itemize}

SRSDD-v1.0: A High-Resolution SAR Rotation Ship Detection Dataset \cite{leiSRSDDv1HighResolutionSAR2021}
\begin{itemize}
  \item Researcher Affiliations: Aerospace Information Research Institute, Chinese Academy of Sciences; Key Laboratory of Technology in Geo-Spatial Information Processing and Application System, Chinese Academy of Sciences; School of Electronic, Electrical and Communication Engineering, University of Chinese Academy of Sciences; Laboratory of Spatial Information Intelligent Processing System, Suzhou Aerospace Information Research Institute; National Key Laboratory of Microwave Imaging Technology, Chinese Academy of Sciences.
  \item Country of Institutions: China.
\end{itemize}

A High Resolution Optical Satellite Image Dataset for Ship Recognition and Some New Baselines \cite{liuHighResolutionOptical2017}
\begin{itemize}
  \item Researcher Affiliations: Institute of Chinese Academy of Sciences; University of Chinese Academy of Sciences; China Academy of Electronics and Information Technology.
  \item Country of Institutions: China.
\end{itemize}

Ships in Satellite Imagery \cite{ShipsSatelliteImagery}
\begin{itemize}
  \item Country of Institutions: United States.
\end{itemize}

Airbus Ship Detection Challenge \cite{AirbusShipDetection}
\begin{itemize}
  \item Country of Institutions: United States.
\end{itemize}

Automatic Ship Classification from Optical Aerial Images with Convolutional Neural Networks \cite{gallegoAutomaticShipClassification2018}
\begin{itemize}
  \item Researcher Affiliations: Pattern Recognition and Artificial Intelligence Group, Department of Software and Computing Systems, University of Alicante; Automation, Robotics and Computer Vision Group, Department of Physics, Systems Engineering and Signal Theory, University of Alicante.
  \item Country of Institutions: Spain.
\end{itemize}

Object Detection in Optical Remote Sensing Images: A Survey and A New Benchmark \cite{liObjectDetectionOptical2020}
\begin{itemize}
  \item Researcher Affiliations: Zhengzhou Institute of Surveying and Mapping; School of Automation, Northwestern Polytechnical University; Department of Cartography, Technical University of Munich.
  \item Country of Institutions: China, Germany.
\end{itemize}

FGSD: A Dataset for Fine-Grained Ship Detection in High Resolution Satellite Images \cite{chenFGSDDatasetFineGrained2020}
\begin{itemize}
  \item Researcher Affiliations: Pattern Recognition and Intelligent System Lab, University of Chinese Academy of Sciences; University of Posts and Telecommunications.
  \item Country of Institutions: China.
\end{itemize}

New Approaches and Tools for Ship Detection in Optical Satellite Imagery \cite{cordovaNewApproachesTools2020}
\begin{itemize}
  \item Researcher Affiliations: Universidad Nacional de San Augustín de Arequipa.
  \item Country of Institutions: Peru.
\end{itemize}

A Public Dataset for Fine-Grained Ship Classification in Optical Remote Sensing Images \cite{diPublicDatasetFineGrained2021}
\begin{itemize}
  \item Researcher Affiliations: Department of Aerospace Information Engineering, School of Astronautics, Beihang University; Beijing Key Laboratory of Digital Media; Key Laboratory of Spacecraft Design Optimization and Dynamic Simulation Technologies, Ministry of Education.
  \item Country of Institutions: China.
\end{itemize}

ShipRSImageNet: A Large-Scale Fine-Grained Dataset for Ship Detection in High Resolution Optical Remote Sensing Images \cite{zhangShipRSImageNetLargeScaleFineGrained2021}
\begin{itemize}
  \item Researcher Affiliations: Department of Electronic Engineering, Tsinghua University; Tsinghua Shenzhen International Graduate School; State Key Laboratory of Space-Ground Integrated Information Technology, China Academy of Space Technology; Aerospace ShenZhou Smart System Technology Company, Ltd.
  \item Country of Institutions: China.
\end{itemize}

Ship Detection in Sentinel 2 Multi-Spectral Images with Self-Supervised Learning \cite{ciocarlanShipDetectionSentinel2021}
\begin{itemize}
  \item Researcher Affiliations: IMT Atlantique; Thales/SIX/ThereSiS.
  \item Country of Institutions: France.
\end{itemize}

LR-TSDet: Towards Tiny Ship Detection in Low-Resolution Remote Sensing Images \cite{wuLRTSDetTinyShip2021}
\begin{itemize}
  \item Researcher Affiliations: Key Laboratory of Technology in Geo-Spatial Information Processing and Application Systems, Institute of Electronics, Chinese Academy of Sciences; Aerospace Information Research Institute, Chinese Academy of Sciences; School of Electronic, Electrical and Communication Engineering, University of Chinese Academy of Sciences.
  \item Country of Institutions: China.
\end{itemize}

VHRShips: An Extensive Benchmark Dataset for Scalable Deep Learning-Based Ship Detection Applications \cite{kizilkayaVHRShipsExtensiveBenchmark2022}
\begin{itemize}
  \item Researcher Affiliations: Satellite Communication and Remote Sensing Program, Informatics Institute, Istanbul Technical University; Geomatics Engineering Department, Civil Engineering Faculty; Istanbul Technical University.
  \item Country of Institutions: Turkey.
\end{itemize}

Accurate Ship Detection Using Electro-Optical Image-Based Satellite on Enhanced Feature and Land Awareness \cite{leeAccurateShipDetection2022}
\begin{itemize}
  \item Researcher Affiliations: School of Electronics Engineering, Kyungpook National University; The Korea Institute of Industrial Technology; The Oceanlightai. Co. Ltd.; Research Center for Neurosurgical Robotic System; School of Mechatronics, Korea University of Technology and Education.
  \item Country of Institutions: Republic of Korea.
\end{itemize}
\end{flushleft}

\end{appendices}


\clearpage
\bibliographystyle{plain}
\bibliography{reference}


\end{document}